\documentstyle[epsfig]{mn}
\newcommand{\etal}{{\it et al.} }        
\newcommand{\mum}{$\mu$m }
\title [Observations of the HDF with ISO - II]
{Observations of the {\it Hubble Deep Field} with the {\it 
Infrared Space Observatory} - II.
Source detection and photometry
\thanks{Based on observations with ISO, an ESA project, with
instruments funded by ESA Member States (especially the PI countries:
France, Germany, the Netherlands and the United Kingdom) 
and with participation of ISAS and NASA.}}
\author[Goldschmidt, Oliver \etal]       
{P. Goldschmidt$^1$, S. Oliver$^1$, S. Serjeant$^1$, A. Baker$^3$, N. Eaton$^1$,
\vspace*{0.2cm}\\{\LARGE  
A. Efstathiou$^1$,
C. Gruppioni$^1$,
R.G. Mann$^1$,
B. Mobasher$^1$,
}\vspace*{0.25cm}\\ {\LARGE 
M. Rowan-Robinson$^1$,
T. Sumner$^1$, 
L. Danese$^2$, 
D. Elbaz$^3$, 
}\vspace*{0.2cm}\\{\LARGE 
A. Franceschini$^4$,
E. Egami$^5$, 
M. Kontizas$^6$, 
A. Lawrence$^7$, 
}\vspace*{0.2cm}\\{\LARGE 
R. McMahon$^8$, 
H.U. Norgaard-Nielsen$^9$, 
I. Perez-Fournon$^{10}$, 
}\vspace*{0.2cm}\\{\LARGE 
I. Gonzalez-Serrano$^{11}$
}\\
$^1$Astrophysics Group, Imperial College London, Blackett Laboratory,
Prince Consort Road, London SW7 2BZ;\\ 
$^2$SISSA, Via Beirut 2-4,Trieste, Italy\\
$^3$Service d'Astrophysique, Saclay, 91191 Gif-sur-Yvette, Cedex, France\\
$^4$Osservatorio Astronomico di Padova, Vicolo dell'Osservatorio 5,
I-35 122, Padova, Italy\\
$^5$Max-Planck-Institut fur Extraterrestrische Physik,
Giessenbachstrasse, D-8046, Garching bei Munchen, Germany\\
$^6$Astronomical Institute, National Observatory of Athens, P.O. Box
200048, GR-118 10,  Athens, Greece\\
$^7$Institute for Astronomy, University of Edinburgh, Blackford Hill,
Edinburgh, EH9 3HJ\\
$^8$Institute of Astronomy, The Observatories, Madingley
Road,Cambridge, CB3 0HA\\
$^9$Danish Space Research Institute, Gl. Lundtoftevej 7, DK-2800
Lyngby, Copenhagen, Denmark\\
$^{10}$Instituto Astronomico de Canarias, Via Lactea, E-38200 La
Laguna, Tenerife, Canary Islands, Spain\\
$^{11}$Instituto de Fisica de Cantabria, Santander, Spain\\
}
\date{Accepted 1997 May 9. Received 1997 March 24; in original form
1996 December 5}
\pagerange{\pageref{firstpage}--\pageref{lastpage}}
\pubyear{1997}
\label{firstpage}

\begin{document}

\maketitle

\begin{abstract}
We present positions and fluxes of point sources found in the Infrared
Space Observatory (ISO) images of the Hubble Deep Field (HDF) at 6.7 and
15 \mum.  
We have constructed  algorithmically selected ``complete'' flux-limited 
samples of 19 sources in the 15 $\mu$m image, and 7 sources in the 6.7 \mum 
image. 
The typical flux limit at 15 \mum is $\sim 0.2$ mJy and 
at 6.7 \mum is $\sim 0.04$ mJy.
We have selected ``supplementary'' samples of 
3 sources at 15 \mum and 20 sources at 6.7 \mum by eye.
We discuss the completeness and reliability of the connected pixel source detection algorithm
used, by comparing the intrinsic and estimated properties of simulated data,
and also by estimating the noise properties of the real data.
The most pessimistic estimate of the number of spurious sources 
in the ``complete'' samples is $1$ at 15 \mum
and $2$ at 6.7~\mum, and in the ``supplementary'' lists is 1 at 15 \mum and
5 at 6.7 \mum.
\end{abstract}


\section{Introduction}

The Hubble Deep Field (HDF, Williams et al., 1996) consists of $\sim 3$ 
square arcminutes near the NGP
that has been surveyed by the HST in 4 wavebands (approximating to $U,B,V,I$) 
to an average limiting magnitude of $m \sim 28$. We have surveyed this area with
ISO (Kessler et al., 1996) using the ISOCAM instrument (Cesarsky et al., 1996)
at 6.7 and 15 \mum.
This paper, the second in a series
which presents and discusses the ISO data, addresses the problem
of source detection and photometry.
Data acquisition and construction of these images are described in 
more detail in paper I by Serjeant et al. (1997). We briefly summarize the steps
taken to reduce the data below.

Section 2 of this paper discusses the results of running the object detection algorithm on 
the data and presents tables of detected objects with their positions and 
fluxes. Section 3 discusses how data were simulated to test the completeness
and reliability of the object detection algorithm and the photometry.

\section{Detected objects and their estimated fluxes}

Paper I (Serjeant et al., 1997) describes the strategy behind the data acquisition and construction
of the images used in this paper. We briefly summarize the main steps
of data acquisition and reduction here.
In order to increase the spatial resolution
of the resulting images, ISO was used in a microscanning mode and rasters 
were made with fractional pixel offsets between each raster centred on each HDF
Wide Field (WF) frame. 
The default dark frame was subtracted from the data which was then
deglitched by identifying and masking out cosmic ray events in the
pixel histories. Flat fields were created by using the mean sky value
measured by each pixel; this method assumes no significant contamination from
genuine objects which would bias the flat field.
Cosmic ray events were filtered out in the time domain, by excluding all
4$\sigma$ events. Longer term effects such as
glitch transients, which could mimic objects, were filtered out at each position
using slightly different techniques at each wavelength. The most obvious way
to exclude glitches is to median filter the data at each pointing, however this
results in lower signal-to-noise than a simple mean. 
It was decided to mosaic the rasters together using
the same ``drizzling'' routine that the optical HDF images were created
with. This meant that the mean of each pointing was taken, then pointings
with integer pixel offsets were medianed together, and then the resulting
medianed images were drizzled together.
The original pixel sizes were 3 and 6 arcseconds at 6.7 \mum
and 15 \mum respectively, after running the drizzling algorithm, the final sizes
were 1 and 3 arcseconds at 6.7 \mum and 15 \mum respectively.

\subsection{The source detection algorithm}
We used the object detection algorithm, PISA, which
counts contiguous pixels above a user-supplied threshold 
(Draper \& Eaton, 1996).
If the number of contiguous pixels exceeds a user-supplied minimum,
then these pixels are defined as being an object.
The minimum number of contiguous pixels that a detected object must have
is set by considering the number of pixels in the point spread function
and also from tests with simulated data. 

PISA uses a constant threshold per pixel across
each image and this threshold depends on the noise, therefore before running PISA  
we split the image up into sub-images, according to the noise
level. We estimated the noise per pixel by looking at the variance from all
the different pixels that had contributed to each pixel in the final drizzled
image. This variance was dominated by Poisson statistics: the area on the sky
that was sampled most by all three overlapping images had the lowest variance,
and the edges which were sampled by the fewest pixels had the highest variance.
As mentioned above, there was so much overlap between the 3 fields making up
the 15 \mum image that we decided to create two sub-images of this image; one
for the edges which had high variance and one for the rest of the image which
had much lower variance.
For the 6.7 \mum image, the overlap region was smaller, and we created 3 sub-images
with different noise values, the central overlap region, the surrounding
region of the individual fields and the edges. 

For the 15 \mum image the detection criteria were as follows: each pixel had
to have a flux greater than $2\sigma$ above the modal sky value in that region,
where $\sigma$ was estimated from the variance in the sky counts and also from
the pixel history. The two estimates of $\sigma$ are in good agreement with each other.
The minimum number of contiguous pixels in a detected object was set to be
8. Running PISA produced 19 objects in the 15 \mum image, using the input
values in table 1.

\subsection{ Photometry of detected sources}

Although PISA estimates total fluxes for detected objects, it was decided to
use a different estimator for the  fluxes of the objects.
The images appear to be confusion limited 
(Paper I), therefore a constant sky value is not a good estimate
of the local background if we are trying to estimate the excess flux due to objects
in faint pixels. 

A better, although not perfect, method is to measure the local
sky, using a concentric annulus around the aperture selected to estimate the
flux of the object. This is still not ideal since
the sky brightness in confusion-limited images fluctuates on spatial scales
similar to the size of the objects. Therefore the sky adjacent to an object
may not be a good estimate of the sky at the position of the object.
A radius of 12 arcseconds for the aperture was chosen because 
this radius encircles $96\%$ of the empirically determined
point spread function (Oliver et al. , 1997, hereafter Paper III).
The sky annulus had an inner radius of 15 arcseconds and an outer radius of
24 arcseconds. Ideally a large sky annulus is preferable if the noise is 
random, but because of the
spatial variations in the sky one needs as ``local'' value for the sky as 
possible.

In the next section  we descibe how we simulated data to estimate errors and
completeness as a function of flux. Here we describe how we used the real 
images to estimate the errors at each wavelength. We created "sky maps"
by convolving the sky annulus with each image to give the estimated sky
value at each pixel. We then subtracted this from the original image and convolved
the resulting image with the object aperture. This image has, at each pixel,
an estimate of the "object-minus-sky" flux that a detected object would have if
it were centred on that image. The variance in this resulting image therefore
gives the variance in the fluxes of the estimated objects. We find that at
15 \mum the 1 sigma error is $\sim 0.1$ mJy and at 6.7 \mum it is $0.02$ mJy.
This is a conservative over-estimate of the noise since we have made no attempt
to mask out genuine objects which presumably will increase the measured 
variance. However these noise estimates do agree with those estimated from
simulated data as discussed below.

\begin{table}
\caption{The sky and threshold intensities in mJy per square
arcseconds used to select objects in the two sub-images of the 15 \mum
image.} 
\begin{tabular}{rrrr}
\hline
Name & Sky  & Threshold \\
Area 1 & 0.402 & $7.6\times 10^{-4}$  \\
Area 2 & 0.402 & $1.6\times 10^{-3}$  \\
\hline
\end{tabular}
\end{table}

\begin{table*}
\caption{Names, positions, estimated fluxes and upper limits at 6.7 \mum 
in mJy , for objects selected from the 15 \mum image by PISA. Objects that
fall outside the 6.7 \mum image are represented by -.}
\begin{tabular}{lccrr}
\hline
Name & RA (2000) & dec (2000) & flux & flux at 6.7 \mum\\
ISOHDF3 J123633.9+621217 & 12 36 33.96 &  +62 12 17.8 &    0.7263  & - \\
ISOHDF3 J123634.3+621238 & 12 36 34.37 &  +62 12 38.6 &    0.4442  & - \\
ISOHDF3 J123635.9+621134 & 12 36 35.95 &  +62 11 34.7 &    0.4196  & - \\
ISOHDF3 J123636.5+621348 & 12 36 36.54 &  +62 13 48.4 &    0.6490  & - \\
ISOHDF3 J123637.5+621109 & 12 36 37.56 &  +62 11 09.6 &    0.2553  & - \\
ISOHDF3 J123639.3+621250 & 12 36 39.33 &  +62 12 50.3 &    0.4333  & $<$0.0975 \\
ISOHDF3 J123641.1+621129 & 12 36 41.11 &  +62 11 29.9 &    0.3763  & $<$0.0895 \\
ISOHDF3 J123643.7+621255 & 12 36 43.73 &  +62 12 55.6 &    0.3186  & $<$0.0432 \\
ISOHDF3 J123646.9+621045 & 12 36 46.98 &  +62 10 45.3 &    0.4119  &  - \\
ISOHDF3 J123648.1+621432 & 12 36 48.13 &  +62 14 32.0 &    0.2310  & 0.0498 \\
ISOHDF3 J123649.8+621319 & 12 36 49.88 &  +62 13 19.9 &    0.4715  & 0.0523 \\
ISOHDF3 J123653.0+621116 & 12 36 53.05 &  +62 11 16.9 &    0.3265  &  - \\
ISOHDF3 J123653.6+621140 & 12 36 53.62 &  +62 11 40.4 &    0.1382  &  - \\
ISOHDF3 J123658.7+621212 & 12 36 58.71 &  +62 12 12.0 &    0.3357  & $<$0.0891 \\
ISOHDF3 J123659.4+621337 & 12 36 59.48 &  +62 13 37.3 &    0.3406  &  - \\
ISOHDF3 J123700.2+621455 & 12 37 00.25 &  +62 14 55.6 &    0.2908  &  - \\
ISOHDF3 J123702.5+621406 & 12 37 02.57 &  +62 14 06.1 &    0.3322  &  - \\
ISOHDF3 J123705.7+621157 & 12 37 05.76 &  +62 11 57.6 &    0.4718  &  - \\
ISOHDF3 J123709.8+621239 & 12 37 09.88 &  +62 12 39.1 &    0.5103  &  - \\
\hline 
\end{tabular}
\end{table*}

\begin{table*}
\caption{Names, positions, estimated fluxes and upper limits at 6.7 \mum
in mJy, for supplementary objects selected
from the 15 \mum image.}
\begin{tabular}{lccrr}
\hline
Name & RA (2000) & dec (2000) & flux & flux at 6.7 \mum \\
ISOHDF3 J123651.5+621357 & 12 36 51.58 &  +62 13 57.2 &  0.1549 & $<$ 0.0514 \\
ISOHDF3 J123658.1+621458 & 12 36 58.12 &  +62 14 58.2 &  0.2104 & - \\
ISOHDF3 J123702.0+621127 & 12 37 02.06 &  +62 11 27.6 &  0.3260 & - \\
\hline 
\end{tabular}
\end{table*}

For each object detected at 15 \mum aperture photometry was performed at the
same position in the 6.7 \mum image in order to determine the flux
at that wavelength. 
The measured flux at 6.7 \mum was designated as a detection
if it were more than the estimated detection limit at that position.
This detection limit was derived by assuming that the objects are not resolved
and therefore, given the PISA criteria one can calculate the total flux of an
object from knowing the point spread function (see Paper III for more details).
Many 15 \mum sources are outside the area of the 6.7 image, and thus
do not have upper limits at 6.7 \mum.
Table 2 lists the positions, fluxes at 15 \mum 
and upper limits at 6.7 \mum estimated for each  object detected
in the 15 \mum field. Section 3 discusses error estimates of the measured
fluxes, using simulated data.

In principle, conversion of the instrumental counts to flux densities is straightforward,
since ISOCAM is a linear device, and involves a constant transformation at all fluxes.
However this assumes that the total on-source integration time is long enough to reach
stabilization. Pre-flight tests suggest that instrumental units 
should be converted to flux densityies in units of mJy by dividing by $M$, where
$M=2.19$ at 6.7 \mum and $M=1.96$ at 15 \mum (ISO-CAM Observer's Manual, 1994).
Subsequently, in-flight data has indicated this value of $M$ should be 
changed to correct for "point-source flux loss". We therefore use $M=1.93$ at 
6.7 \mum and $M=1.57$ at 15 \mum, these are the values recommended by the
CAM instrumental team.
Additionally, the instrument does not always reach
stabilization and typically the first readout measures $\sim 60\%$ of an object's flux. If this
is the case, then $M$ should be further multiplied by $0.6$. 
However, we are not sure that this further correction to $M$ applies 
to our data, and have therefore not used it.
It should be borne in mind that the 
fluxes we present
in tables 2,3,5 and 6 are preliminary.

In addition, a number of objects were 
selected by PISA with slightly less stringent criteria. This
meant that a higher number of spurious objects were selected.
A slightly lower threshold was used in the selection, and the
resulting objects were ``eyeballed'' to see if they looked genuine.
This is because PISA chooses connected pixels regardless of shape, i.e.
a chain of connected pixels above the threshold will be selected as an
object in the same way as a circle of pixels. Eyeballing these
selected objects therefore was designed to select the ``round'' 
objects that appeared to have the similar sort of smooth profiles that one would
expect a genuine object convolved with the point-spread function would have.
These objects will be referred to in this and
subsequent papers as the supplementary objects. Mann et al. (1997, Paper IV) 
discuss the likelihood of these objects being genuine by matching them
up to the optical HDF image. Note that these objects are {\em not} selected
with any reference to the optical images.

Table 3 lists the 
positions and fluxes of these objects. Because the criteria by which these
objects were selected is not fully algorithmic and limiting fluxes cannot be
estimated for them, subsequent papers (e.g. Paper III) which estimate
the number counts do not use these objects. The objects listed in table 2
(and for the 6.7 \mum data, in table 5) will be referred to as the complete
samples.

The 6.7 \mum objects were selected in the same way as the 15 \mum objects.
Aperture photometry was carried out using a radius of 6 arcseconds and a sky
annulus of width 5 arcseconds and inner radius 8 arcseconds.
Table 4 lists the sky levels, and thresholds searched in the 6.7 \mum image.
Table 5 lists the objects detected by PISA in the 6.7 \mum image and table
6 lists the objects in the supplementary catalogue. Again, we measured the
fluxes of these objects at 15 \mum and defined these measurements as being
detections if the flux were larger than the estimated detection limit at that
position.

\begin{table}
\caption{The sky intensities and threshold intensities in mJy per square
arcseconds used to select objects in the three sub-images of the 6.7 \mum
image.}
\begin{tabular}{rrr}
\hline
Name & sky  & threshold  \\
Area 1 & 0.0801 & $9.3\times 10^{-4}$  \\
Area 2 & 0.0802 & $1.1\times 10^{-3}$  \\
Area 3 & 0.0804 & $1.3\times 10^{-3}$  \\
\hline
\end{tabular}
\end{table}

\begin{table*}
\caption{Names, positions, estimated fluxes and upper limits at 15 \mum in mJy, for objects selected
from the 6.7 \mum image by PISA.}
\begin{tabular}{lccrr}
\hline
Name & RA (2000) & dec (2000) & flux & flux at 15 \mum \\
ISOHDF2 J123643.0+621152 & 12 36 43.05 & +62 11 52.9 &    0.0579 & $ < $0.3362\\
ISOHDF2 J123646.4+621406 & 12 36 46.46 & +62 14 06.6 &    0.0521 & $ < $0.3614 \\
ISOHDF2 J123648.2+621427 & 12 36 48.27 & +62 14 27.4 &    0.0657 & $ < $0.2434 \\
ISOHDF2 J123648.4+621215 & 12 36 48.47 & +62 12 15.3 &    0.0512 & $ < $0.2947 \\
ISOHDF2 J123649.7+621315 & 12 36 49.78 & +62 13 15.9 &    0.0481 & 0.4396 \\
ISOHDF2 J123655.1+621423 & 12 36 55.19 & +62 14 23.6 &    0.0304 & $ < $0.2682 \\
ISOHDF2 J123658.8+621313 & 12 36 58.83 & +62 13 13.6 &    0.0674 & $ < $0.2378 \\
\hline 
\end{tabular}
\end{table*}

\begin{table*}
\caption{Names, positions, estimated fluxes and upper limits at 15 \mum 
in mJy, for supplementary objects selected from the 6.7 \mum image.}
\begin{tabular}{lccrr}
\hline
Name & RA (2000) & dec (2000) & flux & flux at 15 \mum \\
ISOHDF2 123641.5+621309 & 12 36 41.57 &  +62 13 09.8 &    0.0213 & $<$0.1311 \\
ISOHDF2 123641.6+621142 & 12 36 41.62 &  +62 11 42.0 &    0.0516 & 0.2646 \\
ISOHDF2 123642.5+621256 & 12 36 42.50 &  +62 12 56.5 &    0.0378 & 0.2062\\
ISOHDF2 123642.6+621210 & 12 36 42.63 &  +62 12 10.9 &    0.0229 & $<$0.2865 \\
ISOHDF2 123642.9+621309 & 12 36 42.95 &  +62 13 09.2 &    0.0511 & $<$0.1700\\
ISOHDF2 123643.1+621203 & 12 36 43.16 &  +62 12 03.6 &    0.0372 & $<$0.2818 \\
ISOHDF2 123643.9+621130 & 12 36 43.93 &  +62 11 30.0 &    0.0504 & $<$0.2252 \\
ISOHDF2 123646.6+621440 & 12 36 46.64 &  +62 14 40.2 &    0.0408 & $<$0.1556 \\
ISOHDF2 123647.1+621426 & 12 36 47.12 &  +62 14 26.5 &    0.0345 & $<$0.2336 \\
ISOHDF2 123648.6+621123 & 12 36 48.69 &  +62 11 23.2 &    0.0371 & $<$0.3924 \\
ISOHDF2 123650.2+621139 & 12 36 50.22 &  +62 11 39.1 &    0.0660 & $<$0.2871 \\
ISOHDF2 123655.2+621413 & 12 36 55.26 &  +62 14 13.9 &    0.0332 & $<$0.2426 \\
ISOHDF2 123655.7+621427 & 12 36 55.71 &  +62 14 27.0 &    0.0396 & $<$0.2675 \\
ISOHDF2 123656.1+621303 & 12 36 56.19 &  +62 13 03.1 &    0.0373 & $<$0.2592 \\
ISOHDF2 123656.6+621307 & 12 36 56.66 &  +62 13 07.3 &    0.0313 & $<$0.2595 \\
ISOHDF2 123657.4+621414 & 12 36 57.40 &  +62 14 14.0 &    0.0381 & $<$0.2434 \\
ISOHDF2 123657.6+621205 & 12 36 57.61 &  +62 12 05.7 &    0.0275 & $<$0.1948 \\
ISOHDF2 123658.6+621309 & 12 36 58.64 &  +62 13 09.3 &    0.0345 & $<$0.2672 \\
ISOHDF2 123658.9+621248 & 12 36 58.93 &  +62 12 48.1 &    0.0431 & $<$0.2788 \\
ISOHDF2 123701.2+621307 & 12 37 01.24 &  +62 13 07.8 &    0.0618 & $<$0.2858 \\
\hline  
\end{tabular}
\end{table*}

\section{Testing the source detection algorithm}

PISA requires that the sky be constant across the image, down to the
noise level used. In other words, we must be sky-noise limited and not
confusion-limited, at least for the brightest pixels in each object that
are used to select that object.
However it is likely that the ISO images of the HDF are confusion-limited
(Paper I; Paper III).
The detectors are sensitive enough to detect faint
objects, but the large telescope
beam means that the flux per object is shared out between many pixels and 
therefore it is difficult to get an unbiased estimate of the background
at the position of the detected objects.

\subsection{Running the source detection algorithm on simulated data}

To see if the source detection algorithm could cope with these data we tested
it by simulating some data and running it through every stage of the source
detection and photometry procedure that was used to construct the ``complete''
samples.

The simulated objects were modelled by assigning flux to individual pixels.
These pixels were then convolved with a gaussian point spread function.
The resulting objects were then laid
down on an image consisting of a patch of ``clean'' sky from the real image.
PISA was then run on this and aperture photometry was performed on the detected
objects, as discussed above.
The criteria by which PISA was judged to have worked were 
(i) the fraction of true objects detected, and 
(ii) the fraction of spurious objects. 
We also estimated the accuracy of the resulting photometry as a function of
true flux.

At 15 \mum PISA detected $70\%$ of the true objects at ${\rm S}=0.225$ mJy, rising to
$92\%$ at ${\rm S}=0.45$ mJy. The mean difference between the estimated and true fluxes
was always consistent with zero, however the fractional
rms scatter in the estimated fluxes for objects with ${\rm S} <0.3$ mJy 
was $\sim 50\%$. For objects with ${\rm S} \ge 0.3$ mJy the fractional rms scatter decreased and
for ${\rm S} \ge 0.4$ mJy the error on the estimated fluxes was of the order of $\sim 30 \%$.
See Fig. 1 for more details of the results of the simulations.
The maximum fraction of spurious objects found by PISA was $\sim 1\%$ (3 spurious
objects) in 10 simulations; in other words, this implies that 0.3 
objects are spurious on the real 15 \mum image.
At 6.7 \mum the fraction of detected objects rose from $68\%$ at $0.04$ mJy to $93\%$ at
$0.065$ mJy. The fractional rms error in the estimated fluxes was $\sim 40 \%$ at ${\rm S} = 0.0425$ mJy
and decreased to $\sim 25 \%$ at ${\rm S} = 0.0625$ mJy. See Fig. 2 for details of the results
of the simulations. The number of spurious objects found was 2 in 7
simulations and so we expect that, again, 0.3 objects detected are spurious.

\begin{figure}
\epsfig{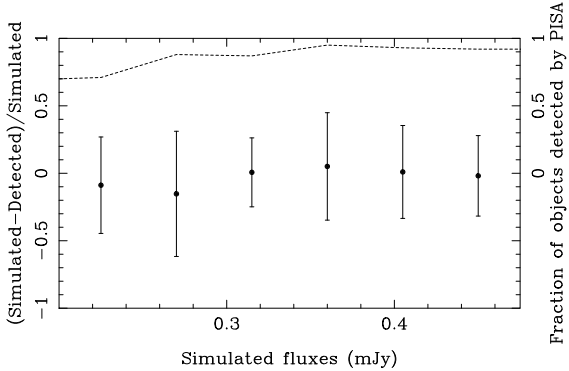}
\caption{The dotted line shows the fraction of simulated objects 
detected by PISA (plotted on the right hand axis)
and the data points show the fractional 
errors in their estimated fluxes at 15 \mum (in mJy, on the left hand axis).}
\label{fig1}
\end{figure}

\begin{figure}
\epsfig{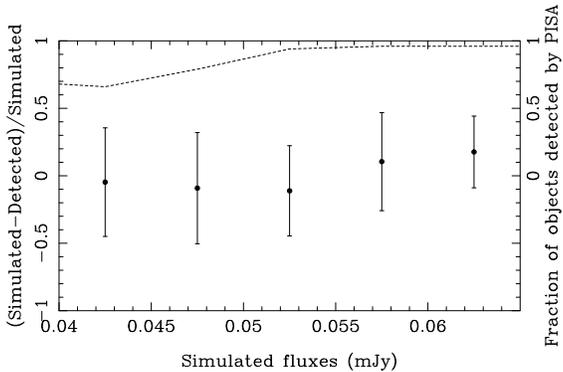}
\caption{The dotted line shows the fraction of simulated objects 
detected by PISA (plotted on the right hand axis)
and the data points show the fractional errors in their estimated fluxes at 6.7 \mum (in mJy, on the left hand axis).}
\label{fig2}
\end{figure}

We can test the PISA algorithm in two other ways. We can look at the
number of `negative' objects that PISA finds, i.e. the number of 
objects which
satisfy the criteria and where each pixel is {\em less} than the mean sky
minus the threshold. This allows us to quantify the number of random fluctuations
which appear as objects. On the 15 \mum image PISA detects 1 negative object,
and on the 6.7 \mum image, PISA detects 2 negative objects. 
However this is not an ideal test as the noise is not likely to be Gaussian.

Secondly, we can use the `noise' maps to search for objects. As discussed in
Serjeant et al. (1997) these maps were created by randomizing the positions of 
the pixels.
Therefore any real objects will be scrambled and will no longer be made
up of contiguous pixels. However noisy pixels which might be spuriously detected
as real objects by PISA will still be present in these noise maps and we can
therefore try and quantify what fraction of our detections are due to noise
by running PISA on the noise maps. PISA finds 2 objects at 6.7 \mum and 1 object at 15 \mum.

To summarize we estimate that, at worst, 1 object at 15 \mum and 2 objects at
6.7 \mum in the ``complete'' lists are spurious detections, and at best,
less than 1 object at each wavelength is spurious.

We then tested likelihood of objects in the `supplementary' sample being
spurious.
We did this by searching for sources on the `noise' maps.
This is more problematic, as it involves human selection `by eye' and thus relies
on this part of the selection being fully reproducible and also unbiassed by
the knowledge that there are no `true' objects on these maps.
A pessimistic estimate is that 1 object at 15 \mum 
and 5 objects at 6.7 \mum in the supplementary lists are spurious.
Mann et al. (1997, paper IV) address this
issue further by matching objects detected in this paper to optically 
selected galaxies in the HDF. 

\section{Summary}

We have used a contiguous-pixel algorithm to detect objects on the ISO images
of the HDF and flanking fields. We have presented tables of objects found at
both 15 \mum and 6.7 \mum. These are the faintest objects ever discovered
at these wavelengths. We estimate from the data that the errors on the measured
fluxes are $\sim 0.1$ mJy ($\sim 30\%$ ) at 15 \mum and $\sim 0.02$ mJy 
($ \sim 40\%$) at 6.7 \mum.
Using simulated data on a real image of the sky we estimate that
down to a flux limit of 0.045 mJy at 6.7 \mum and 0.225 mJy at 15 \mum
PISA can detect $\ge 75\%$ of known objects. 
Also using this simulated data the relative errors in the estimated photometry
of these objects are $\sim 20 - 50 \%$. The simulated data imply that there are
less than one spurious sources at both 15 and 6.7 \mum, whereas using PISA on
the noise maps and counting the number of negative sources implies that 1-2
objects at each of the wavelengths might be spurious in the complete samples.

Further information on the ISO HDF project can be found on the
ISO HDF WWW pages: see http://artemis.ph.ic.ac.uk/hdf/.

\vspace{0.25in}

{\large\bf Acknowledgements}\\

We thank the referee, Leo Metcalfe, for helpful and informative comments.
Data reduction was partially carried out on STARLINK. 
This work was supported by PPARC grant GR/K98728 and by the EC TMR Network programme,
FMRX-CT96-0068.

\label{lastpage}

\begin{thebibliography}{}

\bibitem{} Cesarsky, C., et al., 1996, A \& A, 315, 32
\bibitem{} Draper, P.W. \& Eaton, N. 1996,\\ 
{\tt http://star-www.rl.ac.uk/star/docs/sun109.htx/sun109.html}
\bibitem{} ISO-CAM Observer's Manual version 1.0, 1994, ESA/ESTEC.
\bibitem{} Kessler, M., et al., 1996, A \& A, 315, 27
\bibitem{} Mann, R.G., et al., 1997, MNRAS (Paper IV, this issue)
\bibitem{} Oliver, S., et al., 1997, MNRAS (Paper III, this issue)
\bibitem{} Serjeant, S., et al., 1997, MNRAS (Paper I, this issue)
\bibitem{} Williams, R.E., et al., 1996, AJ, 112, 1335.
\end{thebibliography}
\end{document}